\documentclass[english]{lni}

\IfFileExists{latin1.sty}{\usepackage{latin1}}{\usepackage{isolatin1}}
\usepackage{amsmath}
\usepackage{hyperref}
\usepackage{graphicx}
\usepackage{fancyhdr}
\usepackage{listings} 
\usepackage{changepage} 
\usepackage[figurename=Fig., tablename=Tab., small]{caption}[2008/04/01]

\renewcommand{\headrulewidth}{0.4pt} 
\author{Labib Chowdhury\footnote{Department of Electrical \& Computer Engineering, North South University, Dhaka, Bangladesh, labib.chowdhury@northsouth.edu} \, Mustafa Kamal\footnote{Department of Electrical \& Computer Engineering, North South University, Dhaka, Bangladesh mustafa.kamal@northsouth.edu} \, Najia Hasan\footnote{Department of Electrical \& Computer Engineering, North South University, Dhaka, Bangladesh najia.tasnim@northsouth.edu} \, Nabeel Mohammed\footnote{Department of Electrical \& Computer Engineering, North South University, Dhaka, Bangladesh nabeel.mohammed@northsouth.edu}}
\title{Curricular SincNet: Towards Robust Deep Speaker Recognition by Emphasizing Hard Samples in Latent Space}
\begin{document}

\maketitle

\renewcommand{\refname}{References}
\setcounter{footnote}{2} 

\pagestyle{fancy}
\fancyhead{} 
\fancyhead[RO]{\small CL-SincNet \hspace{25pt}  \hspace{0.05cm}}
\fancyhead[LE]{\hspace{0.05cm}\small  \hspace{25pt} Labib Chowdhury, Mustafa Kamal, Najia Hasan, Nabeel Mohammed}
\fancyfoot{} 
\renewcommand{\headrulewidth}{0.4pt} 

\begin{abstract}
Deep learning models have become an increasingly preferred option for biometric recognition systems, such as speaker recognition. SincNet, a deep neural network architecture, gained popularity in speaker recognition tasks due to its parameterized sinc functions that allow it to work directly on the speech signal. The original SincNet architecture uses the softmax loss, which may not be the most suitable choice for recognition-based tasks. Such loss functions do not impose inter-class margins nor differentiate between easy and hard training samples. Curriculum learning, particularly those leveraging angular margin-based losses, has proven very successful in other biometric applications such as face recognition. The advantage of such a curriculum learning-based techniques is that it will impose inter-class margins as well as taking to account easy and hard samples.  In this paper, we propose Curricular SincNet (CL-SincNet), an improved SincNet model where we use a curricular loss function to train the SincNet architecture. The proposed model is evaluated on multiple datasets using intra-dataset and inter-dataset evaluation protocols. In both settings, the model performs competitively with other previously published work. In the case of inter-dataset testing, it achieves the best overall results with a reduction of  4\% error rate compare to SincNet and other published work.
\end{abstract}

\section{Introduction}
Speaker Recognition(SR) is widely adopted in real-life scenarios as it has brought remarkable changes in security systems, authentication programs, automated identifications, and forensics. SR is divided into two subtasks: - Speaker Verification (SV) and Speaker Identification (SI). SV involves the comparison of two speech signals and determining whether they belong to the same person. It is simply a validation task where the system is required to indicate whether a speech signal given matches the subject who is being considered. Unlike SV, SI is not a validation task but instead can be considered as a search problem, where given a voice sample of a person, the system attempts to identify the speaker from a list of previously registered speakers. 

Before the emergence of deep learning in this field, the popular method included the i-vector method~\cite{ivector}, where the features were extracted from MFCC coefficients and Filter-bank Features~\cite{s13,s17,s18}. These features are then used in a variety of classifiers, including Probabilistic Linear Discriminant Analysis (PLDA)~\cite{plda} and heavy-tailed PLDA~\cite{heavytailed_plda}. Numerous recent SR tasks have been based on the popular SincNet~\cite{sincnet} architecture and as can be appreciated. SR is a very challenging task due to audio signals having a high dimension. Unlike other methods, SincNet can work directly on audio signals because it leverages the parameterized sinc function, which extracts features from audio signals. The deeper layers of the network later process these features.

Biometric systems such as SR and Facial Recognition (FR) can be considered as open-set problems, where the number of classes is not fixed~\cite{allcost}. The original SincNet model was trained using the softmax loss~\cite{sincnet}. Following studies have incorporated various angular margin-based loss functions with SincNet, to achieve better results in both inter-dataset and intra-dataset testing~\cite{amsincnet, allcost}.  While existing models achieve excellent performance on standard datasets~\cite{amsincnet, allcost}, the study performed in~\cite{allcost} demonstrated that these results do not carry over when performing the inter-dataset evaluation, raising a question about the generalizability of these models. To address this, this study proposes the use of a curriculum learning based loss function and incorporates it with the SincNet architecture. Previously curriculum learning based loss function~\cite{curricularface} obtained outstanding performance on biometric tasks such as FR. Influenced by such findings, in this study, we propose Curricular SincNet (CL-SincNet), where we use SincNet architecture as a feature extractor and incorporate curricular loss with it. The contributions of our paper are as follows:
\begin{itemize}
	\item To the best of our knowledge, we are the first one to introduce curriculum learning applied in the angular space to the speaker recognition task.
	\item We conducted extensive experiments on two popular speaker recognition datasets, TIMIT and LibriSpeech, and achieve competitive performance on both. In the case of LibriSpeech, we do better than previously published studies. In fact, our approach reduces the frame error rate by 17\% in intra-dataset testing.
	\item Most significantly, we find our proposed approach achieves a lower Classification Error Rate (CER), compared to previously published models in inter-dataset testing. In fact, our proposed approach reduces the CER by 4\% when trained on LibriSpeech and tested on TIMIT, thus indicating the better generalizability of our approach.
\end{itemize}

\section{Background Study}
This section includes a brief discussion of some background of the softmax loss function and the later part of the discussion includes the SincNet architecture.

Softmax loss is usually defined as the pipeline combination of the last fully connected layer, softmax function, and cross-entropy loss~\cite{amsoftmax}. Softmax loss can be formulated as:
\begin{equation}\label{eqsoftmax}
L_{softmax} = - \frac{1}{N}\sum_{i=1}^{N}log{\frac{e^{W_{k}^{T}f_i + b_{k}}}{\sum_{c=1}^{C} e^{W_{c}^{T}f_{i} + b_{c}}}}
\end{equation}
Here, $f_{i}$ denotes the feature vector from last fully conntected layer, $W_{k}$ represents the $k$th row of weight matrix $W$ and $b_k, b_c$ are the bias scalar value of respective index value $k$ and $c$. $C$ is the total number of classes and the number of training samples in a mini-batch is $N$. By setting bias, $b_{k}, b_{c} = 0$ and ensuring  $W^T_{k}$ and $f_{i}$ are unit norm, equation~\ref{eqsoftmax} can be rewritten equation~\ref{eqnormalized}

\begin{equation}\label{eqnormalized}
L_{softmax}  = - \frac{1}{N}\sum_{i=1}^{N}log{\frac{e^{s \cdot \cos\theta_{k}}}{e^{s \cdot \cos\theta_{k}} + \sum_{c=1,c\neq k}^{C} e^{s \cdot {\cos\theta_c}}}}
\end{equation}

Here $s$ is rescaling parameter and $\theta_{k}$ is angle between weight vector~$W_{k}$ and feature vector~$f_{i}$. Softmax loss in equations~\ref{eqsoftmax}and~\ref{eqnormalized} result in a decision boundary between two classes without having any margin being imposed~\cite{amsoftmax}. However for open-set problems, particularly in biometric recognition areas, margin-based loss functions in particular angular margin-based loss functions have obtained superior and encouraging results~\cite{arcface,curricularface}.

To this end, authors of~\cite{arcface} proposed arcface loss function that mitigates the issue with softmax loss by imposing a margin in angular space, thus creating more robust and larger decision boundaries between classes. The formulation of~\cite{arcface} is as follows:

\begin{equation}\label{eqarcface}
L_{ArcFace} = - \frac{1}{N}\sum_{i=1}^{N}log{\frac{e^{s{\cos(\theta_{{k},i} + m)}}}{e^{s{\cos(\theta_{{k},i} + m)}} + \sum_{c=1,c \neq k}^{C} e^{s {\cos(\theta_{c},i)}}}}
\end{equation}

Where authors added an additional margin with the angle between the target weight vector and the feature vector and then rescale the feature by multiplying with $s$. Although this loss function is verified to obtain good performance~\cite{arcface} it does not consider each sample's difficulties into consideration~\cite{curricularface}. 

The authors of~\cite{curricularface} proposed a new loss function where they leverage curriculum learning and introduce a modulation coefficient in the negative cosine similarity. Authors defined positive cosine similarity as $\cos(\theta_{k} + m)$, which is same as~\cite{arcface} but they changed the representation of negative cosine similarity from $\cos\theta_{j}$ to $N(t,\cos\theta_j)$. The loss is defines as follows:
\begin{equation}\label{eqcurricular}
L_{CurricularLoss} = - \frac{1}{N}\sum_{i=1}^{N}log{\frac{e^{s({\cos(\theta_{{k},i} + m)}}}{e^{s({\cos(\theta_{{k},i} + m)}} + \sum_{c=1,c \neq k}^{C} e^{s {N(t, \cos\theta_{c})}}}}
\end{equation}
The modulation coefficient function $N(t,\cos\theta_c)$ is defined as in~\cite{curricularface}
\begin{equation}\label{eqmodulation}
N(t,\cos\theta_j) = \left\{\begin{matrix}\cos\theta_{j}, \quad \cos(\theta_{k} + m) > \cos\theta_{j}& \\ 
\cos\theta_{j}(t+\cos\theta_{j}), \quad \cos(\theta_{k} + m) < \cos\theta_{j}& 
\end{matrix}\right.
\end{equation}
According to equation~\ref{eqmodulation} a sample is considered to be easy if the angle between the embedding vector and the target weight vector plus the margin is still smaller than the angle between the embedding vector and the weight vector of non-ground truth classes. At the beginning of the training, the hyper-parameter $t$ should be closed to zero so that the model can emphasize the easy samples first, gradually $t$ will increase and the model will focus on the hard example. Since $t$ will increase, the hard sample will be emphasized with larger weights in the later part of the training. the value of $t$ is adaptive in the loss function, as the training goes the estimate of $t$ is formulated as:
\begin{equation}
t^{(k)} = \alpha r^{(k)} + (1-\alpha)t^{(k-1)}
\end{equation}
Here, $r^{(k)}$ is the average of positive cosine similarity of $k$-th batch, defined as $r^{(k)} = \sum_i \cos\theta_{y_{i}}$, $\alpha$ is a momentum parameter, the author from~\cite{curricularface} defined $\alpha = 0.99$.

\section{Proposed model Curricular SincNet}
\begin{figure}[htb]
	\centering
	\includegraphics[scale=1.3]{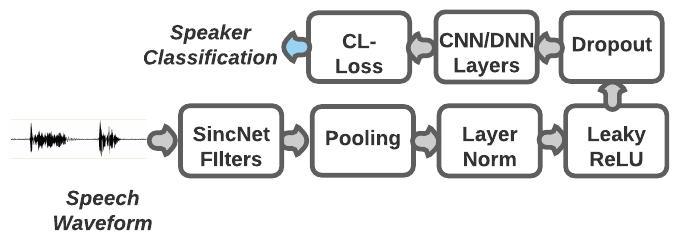}
	\caption{\label{fig:fer} Portrayal of our proposed architecture}
	\label{fig:network}
\end{figure}

Recent work such as~\cite{sincnet, amsincnet, allcost} improved significantly in SI and SV tasks and reports competitive results. Recently~\cite{amsincnet,allcost} used the SincNet architecture and incorporated margin-based losses. The initial convolution operation is performed by using parameterized sinc function to extract low-level features from the raw audio waveform and the only learnable parameters of the convolutional filter are the high and low cutoff frequencies.
The convolution operation is expressed in equation~\ref{eqconvsinc}

\begin{equation}\label{eqconvsinc}
y[n] = x[n] * g[n,\theta]
\end{equation}
Where $x[n]$ represents a chunk of audio/speaker's signal, $y[n]$ is the filtered output. $g$ is a predefined function and it is defined as equation~\ref{eqprefunc}, where $a_{1}$ \& $a_{2}$ represents the low and high cutoff frequencies

\begin{equation}\label{eqprefunc}
g[n,a_1,a_2] = 2a_2\frac{\sin(2\pi a_2n)}{2\pi a_2n} - 2a_1\frac{\sin(2\pi a_1n)}{2\pi a_1n}
\end{equation}

It has been verified that introducing margin-based losses as a last layer of SincNet, helps to increase the distance between classes and decrease the intra-class distance in the embedding space~\cite{allcost,amsincnet}. Although the previously mentioned work showed competitive results, they did not consider each sample's difficulties during the training time. To address this, in this study we propose Curricular SincNet aka CL-SincNet. We use the same feature extractor as others which is SincNet and we use the curricular loss function as outlined in equation~\ref{eqcurricular} to optimize the network. The motivation is that this approach can allow the model to learn features by allowing the network to learn easy samples first and the harder samples later in the training loop. The graphical representation of our CL-SincNet is shown in figure~\ref{fig:network}.

\section{Experimental Setup}

This section describes the details of datasets we use to train and evaluate our model, training and testing procedures, and the metrics we use to measure the model's performance.

\subsection{Datasets}
We consider TIMIT~\cite{timit} and LibriSpeech~\cite{libre} for training and evaluation our model. These two datasets are very popular in SR tasks. TIMIT dataset has 462 classes, or unique speakers, and each class has 8 samples. LibriSpeech has a total 2484 number of classes/unique speakers with a total number of 21933 samples. We used 12-15 seconds of audio for training and 2-6 seconds for testing.

\subsection{Training \& Testing Procedure}\label{sub:metrics}

For the training procedure, we use similar settings as~\cite{sincnet} except for the last layer. We discarded this layer, and instead used the output of the previous layer. We normalized both the feature vector and the row vector of weights by L2 normalization and calculated the cosine similarity of the easy sample and hard sample with corresponding labels. For the two hyperparameters in equation~\ref{eqcurricular} we used the same value as used in~\cite{curricularface}, which is $m = 0.5$ and $s=64$. To train the model we use a mini-batch of $128$ and the learning rate is set to $10^{-2}$.
To evaluate our model, we use the same two protocols as~\cite{allcost} i.e - Intra dataset test and Inter dataset test. An intra-dataset test is performed to evaluate the Speaker Verification performance and an inter-dataset test is performed to evaluate the Speaker Identification performance. 
All codes are available at github's project repository\footnote{ \url{https://github.com/jongli747/Curricular-SincNet}}.

For SV we use Frame Error Rate (FER) and Classification Error Rate (CER) in percentage to demonstrate the performance of our proposed model. These are widely used metrics to measure the performance in SR-based task~\cite{sincnet,amsincnet,allcost}. FER is calculated over a window of 200 ms while CER is calculated by averaging the posterior probabilities computed at each frame of the sample and voting for the speaker with the highest average probability. We also use CER(\%) in inter-dataset testing  for SI task.

The motivation of using the aformentioned metrics is, to demonstrate a fair comparison with recently published works~\cite{allcost, amsincnet, sincnet}.

\section{Results}
In this section, we discuss our results in two parts. Initially, we speak about speaker verification tasks in the intra-dataset protocol, and then we will talk about speaker identification in an inter-dataset protocol.

\subsection{Speaker Verification}

\begin{table}[h]
	\centering
	\begin{tabular}{ccccc}
		\hline
		\multicolumn{1}{}{} &
		\multicolumn{2}{c}{\textbf{FER(\%)}}  & \multicolumn{2}{c}{\textbf{CER(\%)}}  \\
		\hline
		\textbf{Configuration} & \textbf{TIMIT} & \textbf{LibriSpeech}  & \textbf{TIMIT} & \textbf{LibriSpeech} \\
		\hline
		SincNet~\cite{sincnet} & 47.38 & 45.23 & 1.08 & 3.2 \\
		AM-SincNet~\cite{amsincnet} & 28.09 & 44.73 & 0.36 & 6.1 \\
		AF-SincNet~\cite{allcost} & \textbf{26.90} & 44.65 & \textbf{0.28} & 5.7 \\
		Ensemble-SincNet~\cite{allcost} & 35.98 & 45.97 & 0.79 & 7.2 \\
		ALL-SincNet~\cite{allcost} & 36.08 & 45.92 & 0.72 & 6.4 \\
		CL-SincNet (Ours) & 37.36 & \textbf{27.63} & 1.08 & \textbf{0.64} \\
		\hline
	\end{tabular}
	\caption{\label{tab:verification}Comparison of FER(\%) and CER(\%) evaluation for both TIMIT and LibriSpeech}
	
\end{table}

As we mentioned earlier in section~\ref{sub:metrics} we used FER and CER in percentage as evaluation metrics for SV task(intra- dataset test protocol). Table~\ref{tab:verification} presents the FER and CER obtained by our model on the TIMIT and LibriSpeech datasets. The performance reported in the previously published models is also shown for comparison. From table~\ref{tab:verification} we can see that on the LibriSpeech dataset, our proposed model outperforms previously published methods with a significant reduction of FER and CER. In FER, we can see that our proposed model not only outperforms the previously published model, but we have also achieved at least 17\% less error rate on the LibriSpeech dataset. Moreover, in terms of  CER on the LibriSpeech dataset, our proposed approach has achieved the lowest error rate of 0.64\%, reducing the CER by 2.5\% in the speaker verification task. Although our model does not show better performance than~\cite{amsincnet, allcost} on TIMIT, it is worth mentioning that, TIMIT is a comparatively smaller dataset than LibriSpeech. 

\subsection{Speaker Identification (Inter-Dataset Evaluation)}
For the SI task, we usually compare $x:n$, where $x$ is the given speaker's sample and $n$ is a set of registered lists of speakers. Usually, Cosine Similarity or Euclidean Distance is used for evaluation, this study considered Cosine Similarity.

\begin{table}[h]
	\centering
	\begin{tabular}{llll}
		\hline
		\multicolumn{2}{l}{\textbf{Protocol-1}} & \multicolumn{2}{l}{\textbf{Protocol-2}}  \\
		\hline
		\textbf{Configuration} & \textbf{CER (\%)} & \textbf{Configuration}  & \textbf{CER (\%)} \\
		\hline
		SincNet\cite{sincnet} & 10.09 & SincNet\cite{sincnet} & 10.94 \\
		
		AM-SincNet\cite{amsincnet} &  9.39 & AM-SincNet\cite{amsincnet} & 13.10 \\
		
		AF-SincNet\cite{allcost} & 9.14 & AF-SincNet\cite{allcost} & 10.83 \\
		
		Ensemble-SincNet\cite{allcost} & 8.10 & Ensemble-SincNet\cite{allcost} & 12.87 \\
		
		ALL-SincNet\cite{allcost} & 7.15 & ALL-SincNet\cite{allcost} & 10.72 \\
		
		CL-SincNet(Ours) & \textbf{6.39} & CL-SincNet(Ours) & \textbf{6.06} \\
		\hline
	\end{tabular}
	\caption{Comparison of interdataset evaluation for both TIMIT and~LibriSpeech}
	\label{tab:crosstable}
\end{table}
We have adopted the protocol of inter-dataset testing from~\cite{allcost} for the evaluation of the SI task, where the model is trained on one dataset and tested using a different independently collected dataset. This is a good indicator of the generalizability of a model. Table~\ref{tab:crosstable} presents the CER obtained by our model on the TIMIT and LibriSpeech datasets. For the sake of simplicity,  we refer to the TIMIT trained LibriSpeech tested model as protocol-1 and LibriSpeech trained TIMIT tested model as protocol-2. The first two columns represent the result of protocol-1, and the last two columns represent the results of protocol-2. It is worth mentioning that no samples or classes are overlapped between the two datasets. Our proposed CL-SincNet outperforms the previously published model in both settings. At protocol-1, our proposed model has achieved 0.8\% less error rate than compared to previously published work. Most significantly in protocol-2, our proposed CL-SincNet has achieved a $6.06\%$ error rate which is a reduction of more than 4\% error rate than compared to previously published works. As we have mentioned earlier, the TIMIT dataset is small than the LibriSpeech dataset, which may be a reason why an improvement in TIMIT is less significant. Tables~\ref{tab:verification},~\ref{tab:crosstable} indicate that our proposed CL-SincNet has the capacity to generalized better than other published models with significant performance improvements.

\section{Conclusions}

This study has proposed Curricular SincNet (CL-SincNet), where we leverage an angular margin-based curriculum learning loss function on the SincNet architecture for the speaker recognition task. The proposed CL-SincNet has manifested superior results compared to previously published studies~\cite{sincnet, amsincnet,allcost}. Our proposed approach reduces the frame error rate by 17\% on the LibriSpeech dataset for speaker verification tasks in intra-dataset test protocol and reduces the 4\% classification error rate in inter-dataset testing for speaker identification tasks. The results indicate that introducing such a curriculum learning-based loss function on SincNet architecture can have positive outcomes for open-set biometric recognition systems. 

\bibliographystyle{lnig}
\bibliography{lniguide_en}

\end{document}